\documentclass[aps,preprint,floatfix,showpacs]{revtex4-1}
\usepackage{epsf}
\usepackage{epsfig}
\usepackage{color}
\usepackage{graphicx}
\usepackage{amssymb}
\usepackage{amsmath}
\usepackage{color}
\usepackage{float}
\usepackage{comment}
\usepackage{setspace}
\usepackage{epstopdf}
\usepackage{subfigure}
\usepackage{setspace}
\definecolor{dark red}{rgb}{0.75, 0.0, 0.0}

\begin{document}

\title{Site-disorder driven superconductor-insulator transition: A dynamical mean field study}
\author{Naushad Ahmad Kamar and N.\ S.\ Vidhyadhiraja}
\affiliation{Theoretical Sciences Unit\\Jawaharlal Nehru Centre for Advanced Scientific Research,\\
 Jakkur, Bangalore 560 064, India.}

\begin{abstract}
We investigate the effect of site-disorder on the superconducting (SC) state in the
attractive Hubbard model within the framework of dynamical mean field theory.
For a fixed interaction strength ($U$), the SC order parameter (OP) decreases monotonically with increasing disorder ($x$), while the single-particle spectral gap (SG) decreases for small $x$,
reaches a minimum and keeps increasing for larger $x$. Thus, the system remains gapped beyond the destruction of the superconducting state, indicating a disorder-driven superconductor-insulator transition. We investigate this transition
in depth considering the effects of weak and strong disorder for a range of 
interaction strengths. 
In the clean case, the order-parameter is known to increase monotonically with increasing interaction, saturating
at a finite value asymptotically for $U\rightarrow \infty$.
The presence of disorder results in destruction of superconductivity at large $U$, thus
drastically modifying the clean case behaviour. A physical understanding of our findings is obtained by invoking particle-hole asymmetry and the probability distributions of the order parameter and spectral gap.
\end{abstract}


\maketitle

\section{Introduction}
The combined effect of disorder and correlations on the superconducting state has been
extensively studied since many decades but a complete picture has not emerged yet~\cite{PA,Belitz}.  
Many recent experimental studies of disordered superconducting thin films have investigated the superconductor-insulator
 transition (SIT)~\cite{Pratap1, Pratap2, Crane}. In disordered NbN s-wave superconductors~\cite{Pratap1, Pratap2}, an SIT was observed through scanning tunneling spectroscopy and penetration depth measurements. The effective disorder in NbN, given by the product of Fermi wave vector
 ($k_F$) and electronic mean free path ($l$), is introduced by controlling the vacancy of Nb atoms. The insulating state at large disorder has been a subject of debate. Finite frequency measurements of superfluid stiffness~\cite{Crane}
have indicated the existence of a Bose insulator state with localized Cooper pairs. However,
a better understanding of the SIT and the associated insulating state in the large disorder
limit requires more theoretical scrutiny.

    Localized Cooper pairs indicate a system with local attractive interactions, which
are most appropriately represented by the attractive Hubbard model (AHM). The
study of dirty superconductors can thus be naturally carried out by investigating the effect of disorder in the attractive Hubbard model.  The clean limit of the AHM has been
extensively studied using Bogoliubov-de Gennes type mean field (BdGMF) theories and more recently using iterated perturbation theory with superconducting bath (IPTSC)~\cite{Arti}, numerical renormalization group (NRG)~\cite{Bauer} and continuous time
quantum Monte Carlo (CTQMC)~\cite{Werner} within dynamical mean field theory (DMFT)~\cite{Antoine, Kotliar, Dieter}. The main
issue that has been focused on is the BCS-BEC crossover for different fillings
and interaction strengths. Very recently, a statistical-DMFT study of the 
Bethe lattice disordered AHM has been carried out~\cite{Kawa}. 

The systems investigated experimentally, such as NbN$_x$~\cite{Pratap1, Pratap2} and InO$_x$~\cite{Crane} are in the 
strong disorder limit, wherein the
early theories of dirty superconductors, e.g by Anderson~\cite{Anderson} and by Abrikosov and Gorkov~\cite{Gorkov}
are not really applicable. Hence a proper theoretical approach is needed which treats the effects
of strong attractive interactions and disorder on an equal footing.  One such method,
namely the BdGMF
has been employed to investigate the AHM with site\cite{Amit} as well as bond disorder~\cite{Sanjeev}. However,
the method, being based on a mean field approximation, albeit inhomogeneous, has
limitations in terms of not incorporating quantum fluctuations. Recent NRG~\cite{Bauer} calculations
of the clean AHM also point out several deficiencies of the BdGMF method. Beyond mean field,  QMC~\cite{Nandini} studies of finite size lattices have validated the decrease of superconducting order
parameter ($\Phi$) with increasing disorder, but due to finite size effects, a complete destruction of $\Phi$ at large disorder could not be seen. However, with increasing temperature, a superconductor-insulator transition was observed in the dc conductivity 
calculations. 

In this work, we carry out
a detailed study of the disordered AHM
 by combining coherent potential approximation(CPA)~\cite{Elliott, Jannis}, with DMFT and iterated perturbation theory for superconductivity(IPTSC).
The IPTSC
solver has the advantage over methods such as QMC of obtaining real frequency spectra at zero temperature and in the thermodynamic limit, while being computationally inexpensive. The reliability of our approach is enhanced by the fact that the IPTSC is known to benchmark well with NRG results for the clean AHM. To distinguish between
dynamical and static effects, we have also carried out BdGMF studies within CPA+DMFT.
As anticipated 
by the previous inhomogeneous mean field and QMC calculations, we find a SIT with increasing disorder. We map out the detailed behaviour of the SIT in the disorder-interaction plane. 
We also investigate the distribution of the local order parameter, and point out that some of the subtle aspects
 of the experimentally observed SIT require an extension of single-site DMFT through
e.g the statistical DMFT or cluster extensions such as the dynamical cluster approximation.
The paper is structured as
follows: In the following section, we outline the model and the formalism used. Next, we 
present our results for the local order parameter, the spectra, the distribution of the
order parameter and the phase diagram. We conclude in the final section.

\section{Model and formalism}
 The attractive Hubbard model (AHM) is expressed, in standard second quantized
notation by the following Hamiltonian:
\begin{equation}
{\cal H}=
\sum_{i\sigma}\epsilon_i c^{\dagger}_{i\sigma}c^{\phantom{\dagger}}_{i\sigma}-t\sum_{\langle ij\sigma\rangle} \left(c^{\dagger}_{i\sigma}c^{\phantom{\dagger}}_{j\sigma}+h.c\right)-|U|\sum_{i} 
n_{i\uparrow}n_{i\downarrow }-
\mu\sum_{i\sigma}c^{\dagger}_{i\sigma}c^{\phantom{\dagger}}_{i\sigma}
\label{eq:eq1}
\end{equation}
where ${c}_{i\sigma}$ annihilates an electron on $i^{\rm th}$ lattice site with spin $\sigma$, and ${n}_{i\sigma}=c^{\dagger}_{i\sigma}c^{\phantom{\dagger}}_{i\sigma}$;
$t$ is nearest-neighbour hopping amplitude, ${\epsilon}_i$ is site-energy, and $\mu$ is chemical potential. The disorder is represented by randomness in  site energies, which 
we choose to be distributed according to a uniform probability distribution function $P_\epsilon(\epsilon_{i})$,
\begin{equation}
{P_\epsilon(\epsilon_{i})}=\frac{\Theta(\frac{x}{2}-|\epsilon_i|)}{x}
\label{eq:eq2}
\end{equation}
where $x$ is  disorder strength in unit of $t \,(=1.0)$.

The CPA in conjunction with DMFT is the best single-site approach to study
the interplay of disorder with interactions in strongly correlated systems~\cite{pott}.
 To investigate the effects of disorder on the superconducting state, we employ 
the best single-site quantum approaches, namely DMFT in conjunction with CPA. 
Within DMFT, the lattice model is mapped onto a single-impurity model
embedded in a self-consistently determined bath. For the present problem,
the effective medium is in a superconducting state, hence the Nambu
formalism must be used.
The impurity Green's function in Nambu formalism is given as
 \begin{equation} \label{eq:eq3} \hat{G^{i}}^{-1}(\omega^{+}) =
\begin{pmatrix}
  \omega^{+}+\mu-\epsilon_{i}-\Delta_{11}(\omega^{+})-\Sigma_{i}(\omega^{+}) & -\Delta_{12}(\omega^{+})-S_{i}(\omega^{+})\\
 -\Delta_{21}(\omega^{+})-S_{i}(\omega^{+}) & \omega^{+}-\mu+\epsilon_{i}-\Delta_{22}(\omega^{+})+\Sigma^{*}_{i}(-\omega^{+})
\end{pmatrix}
\end{equation}
where $\Delta_{\alpha\beta},\, \alpha,\beta=1,2$ are components of the disorder-averaged hybridisation function matrix $\hat\Delta$, $\Sigma_{i}$ and $S_{i}$ are 
normal and anomalous self-energies of the $i^{\rm th}$ site  respectively.

To calculate the local self-energies, $\Sigma_{i}(\omega^{+})$ and 
$S_{i}(\omega^{+})$ for the $i^{\rm th}$ site,
we use  iterated perturbation theory for superconductivity (IPTSC)\cite{Arti} as
 the impurity solver. In the IPTSC method, based on second order perturbation theory,
the self-energies are given by the following ansatz:
\begin{eqnarray}
\Sigma_{i}(\omega^{+})&=&-U\frac{n_i}{2}+A_i\Sigma^{(2)}_{i}(\omega^{+})
\label{eq:eq4} \\
S_{i}(\omega^{+})&=&-U\Phi_i + A_iS^{(2)}_{i}(\omega^{+})\,
\label{eq:eq5}
\end{eqnarray}
where the local filling $n_i$ and order parameter $\Phi_i$ are given by
\begin{eqnarray}
n_i&=&-\frac{2}{\pi}{\rm Im}\int_{-\infty}^{\infty}d\omega\,G^{i}_{11}(\omega^+)\,f(\omega)
\label{eq:eq6}\\
\Phi_i&=&\int_{-\infty}^{\infty}d\omega\frac{-{\rm Im}{G^i}_{12}(\omega^+)}{\pi}
f(\omega)
\label{eq:eq7}
\end{eqnarray}
and $f(\omega)=\theta(-\omega)$ is the Fermi-Dirac distribution function at zero temperature.
In the ansatz above (equations~(\ref{eq:eq4},\ref{eq:eq5})), the second order self-energies are given by
\begin{eqnarray}
\Sigma^{(2)}_{i}(\omega^{+})&=&U^2\int_{-\infty}^{\infty}\prod_{j=1}^{3}d\omega_{j}\frac{g_{1i}(\omega_1,\omega_2,\omega_3)N(\omega_1,\omega_2,\omega_3)}{\omega^{+}-\omega_1+\omega_2-\omega_3} \nonumber \\
{\mbox{and}} \nonumber \\
S^{(2)}_{i}(\omega^{+})&=&U^2\int_{-\infty}^{\infty}\prod_{j=1}^{3}d\omega_{j}\frac{g_{2i}(\omega_1,\omega_2,\omega_3)N(\omega_1,\omega_2,\omega_3)}{\omega^{+}-\omega_1+\omega_2-\omega_3} \,,
\label{eq:eq8}
\end{eqnarray}
where 
\begin{eqnarray}
N(\omega_1,\omega_2,\omega_3)&=&f(\omega_1)f(-\omega_2)f(\omega_3)+f(-\omega_1)f(\omega_2)
f(-\omega_3) \nonumber \\
g_{1i}(\omega_1,\omega_2,\omega_3)&=&\tilde{\rho}^i_{11}(\omega_1)\tilde{\rho}^i_{22}(\omega_2)\tilde{\rho}^i_{22}(\omega_3)-\tilde{\rho}^i_{12}(\omega_1)\tilde{\rho}^i_{22}(\omega_2)\tilde{\rho}^i_{12}(\omega_3)\nonumber \\
g_{2i}(\omega_1,\omega_2,\omega_3)&=&\tilde{\rho}^i_{12}(\omega_1)\tilde{\rho}^i_{12}(\omega_2)\tilde{\rho}^i_{12}(\omega_3)-\tilde{\rho}^i_{11}(\omega_1)\tilde{\rho}^i_{12}(\omega_2)\tilde{\rho}^i_{22}(\omega_3)
\label{eq:eq9}
\end{eqnarray}
and the spectral functions $\tilde{\rho}^i_{\alpha\beta}$, $\alpha,\beta=1,2$ are given by the imaginary part
of the `Hartree-corrected' host Green's function, namely
${\hat{\tilde{\rho_i}}}=-{\rm Im}\hat{\cal {G}}^i/\pi$. The latter is given by  
\begin{equation} \label{eq:eq10} \hat{\cal{G}}^i(\omega^{+})^{-1} =
\begin{pmatrix}
  \omega^{+}+\mu-\epsilon_i-\Delta_{11}(\omega^{+})+U\frac{n_i}{2} &-\Delta_{12}(\omega^{+})+U\Phi_i\\
 -\Delta_{21}(\omega^{+})+U\Phi_i & \omega^{+}-\mu+\epsilon_i-\Delta_{22}(\omega^{+})-U\frac{n_i}{2}
\end{pmatrix}\,.
\end{equation}
Finally the coefficient $A_i$ in the IPTSC ansatz (equation~(\ref{eq:eq4},{\ref{eq:eq5}}), which
is determined by the high frequency limit is given by
\begin{equation}
A_i=\frac{\frac{n_i}{2}(1-\frac{n_i}{2})-\Phi_i^2}{{\frac{n_{0i}}{2}(1-\frac{n_{0i}}{2})-\Phi_{0i}^2}}\,,
\label{eq:eq11}
\end{equation}
where the pseudo order-parameter $\Phi_{0i}$ and the pseudo occupancy $n_{0i}$ are given by 
\begin{eqnarray}
n_{0i}&=&2\int_{-\infty}^{\infty}d\omega \tilde{\rho}^i_{11}(\omega) f(\omega) \nonumber \\
{\mbox {and}}\;\;\Phi_{0i}&=&\int_{-\infty}^{\infty}d\omega \tilde{\rho}^i_{12}(\omega) f(\omega)
\label{eq:eq12}
\end{eqnarray}

Using the coherent potential approximation (CPA) for incorporating disorder, the
 CPA Green's function is given by an arithmetic averaging over local Green's functions as
\begin{equation}
\hat{G}^{CPA}(\omega^{+})=\int_{\frac{-x}{2}}^{\frac{x}{2}} d{\epsilon_{i}}\hat{G^{i}}(\omega^{+})P_{\epsilon}({\epsilon_{i}})\,.
\label{eq:eq13}
\end{equation}
Since the CPA maps the disordered problem onto a translationally invariant problem, a
lattice Green's function may then be defined as 
 \begin{equation} \label{eq:eq14} \hat{G}_{latt}{(\vec{k},\omega^{+})}^{-1} =
\begin{pmatrix}
\omega^{+}+\mu-\epsilon{(\vec{k})}-\Sigma_{11}^{CPA}(\omega^{+}) &-\Sigma_{12}^{CPA}(\omega^{+})\\
 -\Sigma_{21}^{CPA}(\omega^{+}) & \omega^{+}-\mu+\epsilon{(\vec{k})}-\Sigma_{22}^{CPA}(\omega^{+})
\end{pmatrix}
\end{equation}
where the self-consistency condition is that the lattice self-energy is the same
as the CPA self-energy, hence the
CPA Green's function in term of average self-energy is given as
 \begin{equation} \label{eq:eq15} \hat{G}{^{CPA}}(\omega^+)^{-1} =
\begin{pmatrix}
  \omega^{+}+\mu-\Delta_{11}(\omega^{+})-\Sigma_{11}^{CPA}(\omega^{+}) & -\Delta_{12}(\omega^{+})-\Sigma_{12}^{CPA}(\omega^{+})\\
 -\Delta_{21}(\omega^{+})-\Sigma_{21}^{CPA}(\omega^{+}) & \omega^{+}-\mu-\Delta_{22}(\omega^{+})-\Sigma_{22}^{CPA}(\omega^{+})
\end{pmatrix}
\end{equation}
The equations are closed by observing that the $\vec{k}$ summed lattice
Green's function should be the CPA Green's function, i.e
\begin{eqnarray}
\frac{1}{N_{s}}\sum_{\vec{k}}\hat{G}_{latt}{(\vec{k},\omega^{+})}=\hat{G}^{CPA}(\omega^{+}) \nonumber \\
\int_{-\infty}^{\infty}d\epsilon\rho_{0}(\epsilon) \hat{G}_{latt}{({\epsilon},\omega^{+})}=\hat{G}^{CPA}(\omega^{+})
\end{eqnarray}
where $N_{s}$ and $\rho_{0}(\epsilon)$ are number of lattice sites and non-interacting density of state respectively.
In practice, we follow the steps outlined below to obtain the converged order parameter and spectra.
\begin{enumerate}
\item Guess a hybridisation matrix $\hat\Delta(\omega^+)$ and  $n_i, \Phi_i$ for each site. In practice, we choose either a previously converged solution or the non-interacting
$\hat\Delta(\omega^+)$ with $n_i=1$ and $\Phi_i=1/2\;\forall\; i$.

\item Given a hybridization, occupancy and order parameter, use equation (\ref{eq:eq10}) to calculate the host Green's function matrix, ${\cal{\hat G}}^i(\omega^+)$.

\item From equation (\ref{eq:eq12}) , calculate pseudo-occupancy and pseudo-order parameter, $n_{0i}$ and $\Phi_{0i}$.

\item Now by using equations~(\ref{eq:eq4}, \ref{eq:eq5}, \ref{eq:eq8}, \ref{eq:eq9}, \ref{eq:eq11}) calculate the regular and anomalous self-energies, $\Sigma_i(\omega^+)$ and $S_i(\omega^+)$.

\item Then by using equations~(\ref{eq:eq3}, \ref{eq:eq6}, \ref{eq:eq7}), calculate impurity Green's function $\hat G^i(\omega^+)$, $n_i,\Phi_i$  for each site.

\item The disorder-averaged Green's function, $\hat{G}^{CPA}(\omega^+)$ is obtained using equation~(\ref{eq:eq13}).

\item We consider the AHM on Bethe lattice of infinite connectivity at half filling, which is 
achieved by setting $\mu=-U/2$. For Bethe lattice the self-consistency condition is simply given by
\begin{equation}
\hat{\Delta}(\omega^+)=\frac{t^2\sigma_{z}\hat{G}^{CPA}(\omega^+)\sigma_{z}}{4}
\label{eq:eq17}
\end{equation}
where $\sigma_{z}$ is z component of Pauli's matrix. Using equation~(\ref{eq:eq17}), a new hybridisation matrix $\hat\Delta$ is obtained. 

\item If the hybridisation matrix $\hat\Delta(\omega^+)$ from step 7 and $n_i,\Phi_i$ from step 5 are equal (within a desired numerical tolerance) to the guess  hybridisation matrix $\hat\Delta(\omega^+)$, $n_i$ and $\Phi_i$ from step 1, then the iterations may be stopped, else the iterations continue until desired convergence is achieved.

\end{enumerate}

The results obtained using the above-mentioned procedure will be denoted as IPTSC.
We have also carried out mean-field calculations by `turning off' the dynamical 
self-energies in equations~\ref{eq:eq4} and ~\ref{eq:eq5}. These results will be denoted as BdGMF. We present our results in the next section.

\section{Results and discussion}
  A recent study of the
clean AHM ($x=0$) has shown that results obtained using NRG~\cite{Bauer}
compare well with  those from the IPTSC method, thus indicating its reliability
for the present problem. A total of
1600 lattice sites have been considered in our calculations. This implies that
the impurity problem needs to be solved 1600 times for each DMFT iteration.
A fully parallel implementation allows us to carry out efficient calculations
in a wide-parameter range. This must be contrasted with previous state-of-the-art QMC calculations which have been carried out on a $8\times 8$ square lattice, and results have been obtained at finite temperature on the imaginary frequency Matsubara axis.  Thus not only are we able to consider much larger lattice sizes than previous works, but also obtain real frequency spectra directly at zero temperature.

We review the physics of the clean AHM within DMFT briefly. At half-filling ($n=1$), it is well known that
the AHM has two instabilities, namely superconductivity and charge-density
wave (CDW). If the CDW instability is ignored, then the superconducting order parameter ($\Phi$)
becomes non-zero only for $U> U_{c1}$. The ground state is a normal metal
for low interactions, while for $U>U_{c1}$, the single-particle spectrum develops
a BCS superconducting gap. It is known from NRG calculations, that agree very well with IPTSC
results, that with increasing $U$, the order parameter increases
and saturates to a finite value as $U\rightarrow \infty$. However, this result does not agree with more recent 
CTQMC based calculations~\cite{Werner}, which show that at large $U$, the order parameter decreases again, and vanishes 
beyond a certain $U=U_{c2}$. The contradicting results from NRG and CTQMC need further scrutiny.
In this work, we will choose to take the NRG results as our benchmark.
 
\subsection{Varying disorder; fixed interaction strength}

In the clean case, the half-filling condition is maintained by choosing $\epsilon_i=0$
and $\mu=-U/2$. For $x>0$ the half-filling condition is again maintained
through $\mu=-U/2$. The individual sites have site-energies that
are distributed uniformly over $[-x/2,x/2]$, so there is very little probability
that any single-site would have exactly $\epsilon_i=0$. This implies that for
a disordered AHM at a global half-filling condition, the individual sites
are away from half-filling, hence the CDW instability need not be considered. 
\begin{figure}[ht]
\centering
\includegraphics[scale=0.4,clip]{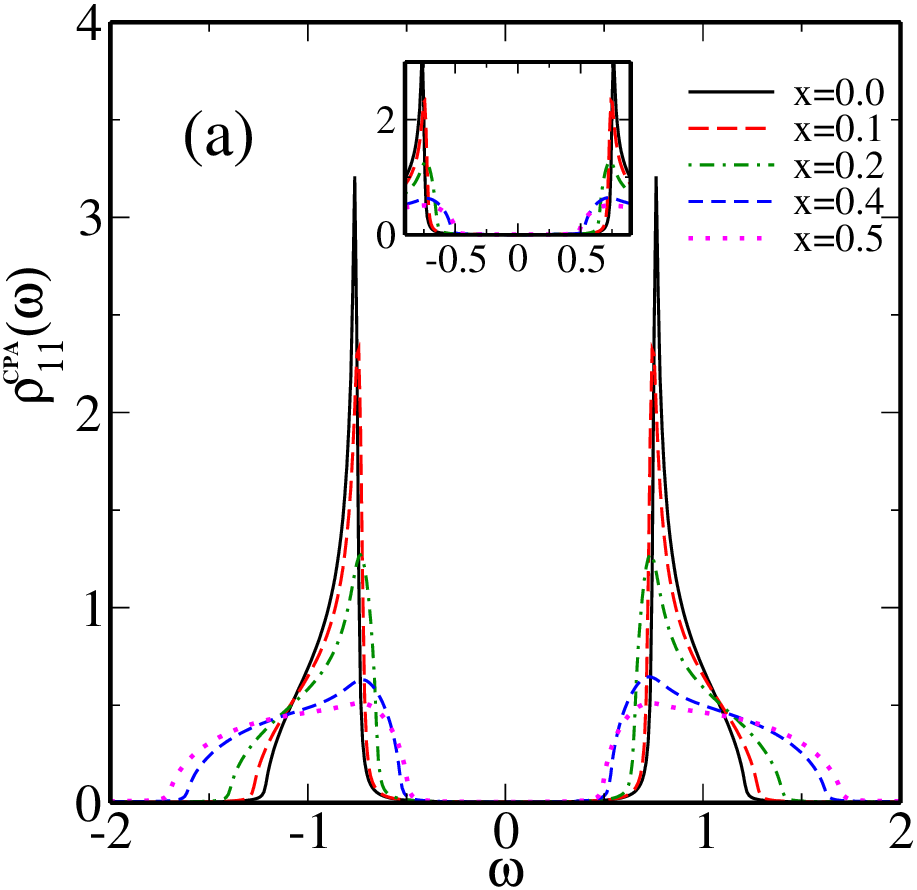}
\includegraphics[scale=0.4,clip]{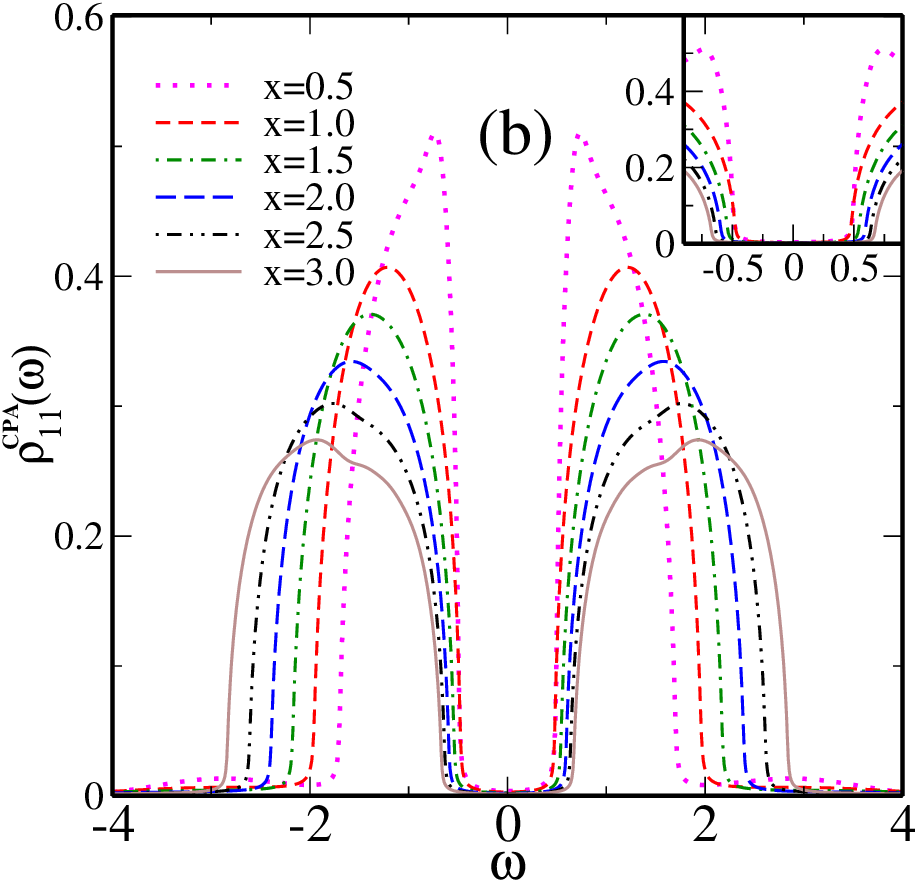}
\caption{(color online) Diagonal spectral function as function of frequency for different value of x at U=2.0: (a) low disorder results ($0<x\leq 0.5$); (b) strong disorder results  ($x\geq 0.5$). Insets show an expanded view of the low frequency gap region. }
\label{fig:fig1}
\end{figure}
In Fig.~\ref{fig:fig1}  we show the diagonal spectral function as a function of frequency
 for different values of disorder at a fixed interaction strength, namely $U=2$. The 
panel (a)
 represents results for lower disorder ($x\le   
 0.5$), while the panel (b) represents higher disorder ($x\ge 0.5$). The zero disorder
case in the panel (a) represents the clean AHM result. The system has a superconducting
 gap $E_g$. Flanking the band edges are the sharp `coherence peaks'.  With
increasing disorder, the coherence peaks melt and change into broad  features. The insets show
 an expanded view of the low frequency gap region.

\begin{figure}[ht]
\centering
\includegraphics[scale=0.4,clip]{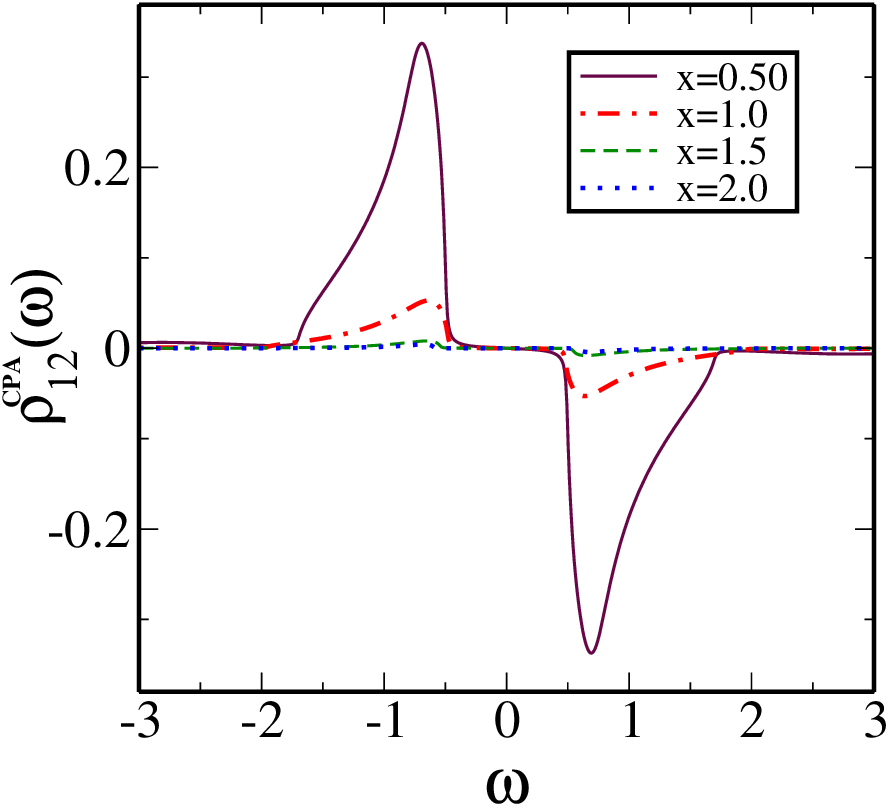}
\caption{Off diagonal spectral function for different values
 of $x$ at a fixed $U=2.0$.}
\label{fig:fig2}
\end{figure}
To understand the nature of the gap, we need to analyse the off-diagonal spectral
function. These are shown in Fig.~\ref{fig:fig2}, for disorder values varying
from $x=0.5$ to $x=2$, and $U=2$. It is seen clearly that the entire spectral weight
decreases rapidly with increasing $x$, and finally goes to zero at $x\sim 1.5$.
Comparing with Fig.~\ref{fig:fig1}, we see that the spectral gap does not close
for any $x$. Thus, the system exhibits a superconductor-insulator transition (SIT)
as a function of increasing disorder at a fixed $U$. 

The effective averaged s-wave pairing amplitude, $\Phi^{CPA}$ is defined as
\begin{equation}
\Phi^{CPA}=\int_{-\infty}^{\infty}d\omega\frac{-{\rm Im}(G^{CPA}_{12}(\omega^+))f(\omega)}{\pi}\,,
\end{equation}
and computed using the mean field (dashed line) and IPTSC (solid line) methods
for a fixed interaction strength. 
\begin{figure}[ht]
\centering
\includegraphics[scale=0.4,clip]{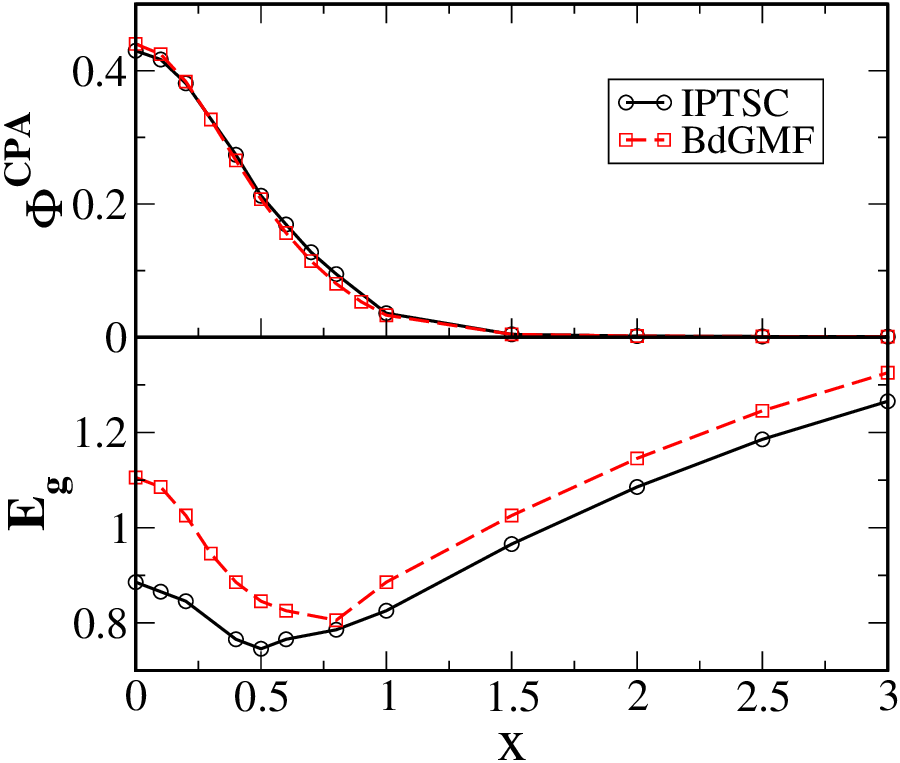}
\caption{Top panel: The disorder-averaged superconducting order parameter (solid line) as a function of $x$ at $U=2.0$. The dashed line
is the BdGMF result for comparison. Bottom panel: The spectral gap as a function of disorder.}
\label{fig:fig3}
\end{figure}
In Fig.~\ref{fig:fig3}, we show the spectral gap
and the disorder-averaged superconducting order parameter 
as a function of disorder in the bottom and top panels respectively.
It is observed that the gap decreases, reaches a minimum, and then increases
with increasing disorder. This kind of behaviour of spectral gap with
disorder is also reported in\cite{Amit,Sanjeev}. The order parameter, in contrast to the gap,
decreases monotonically with increasing disorder and vanishes beyond $x\sim 1.5$. 
This result, which states that the superconducting state is destroyed beyond a critical disorder strength
is consistent with previous QMC results, although the latter were obtained through extrapolation for 
finite size lattices.
We show the BdGMF(CPA) result (dashed line) also in the same figure, and it is seen that the mean-field result and the full DMFT
result hardly differ, indicating that local quantum fluctuations are not playing a significant role in the destruction
of the superconducting state.
Since the order parameter is finite for $x \lesssim 1.5$,
the spectral gap is a superconducting gap, while for higher disorder ($x>1.5$), since $\Phi=0$,
$E_g$ represents an insulating gap. 
We also show the BdGMF(CPA) results for $\Phi$
and $E_g$ in the
same figure. It is seen that the spectral gap is overestimated by the static mean field approach for finite
disorder, as was found at $x=0$~\cite{Arti}.

The dependence of the gap on $x$ is non-monotonic,
and deserves some attention. For weak disorder, the gap decreases with increasing $x$, and this
may be understood through the clean AHM. For $x=0$, the spectral gap becomes smaller
and asymmetric (about the chemical potential),
with either increasing or decreasing the filling away from 1~\cite{Werner}. Thus, we expect that
with increasing disorder, since most sites would be off-half-filling, the spectral gap
within CPA, arising as the arithmetic mean of the individual spectra, would
decrease for weak disorder. The preceding argument assumes that the hybridization remains
largely unaffected, which is true for weak disorder. However, for moderate and large disorder, the
 hybridization gets modified strongly, and the simple arithmetic averaging result cannot be used to understand the
increase in gap at larger disorder values ($x\gtrsim 1.5$). In order to understand the
behaviour at large disorder, we will need to probe the distribution of the order
parameter and gap over all the sites in the lattice. This is considered next.

\begin{figure}[ht]
\centering
\includegraphics[scale=0.4,clip]{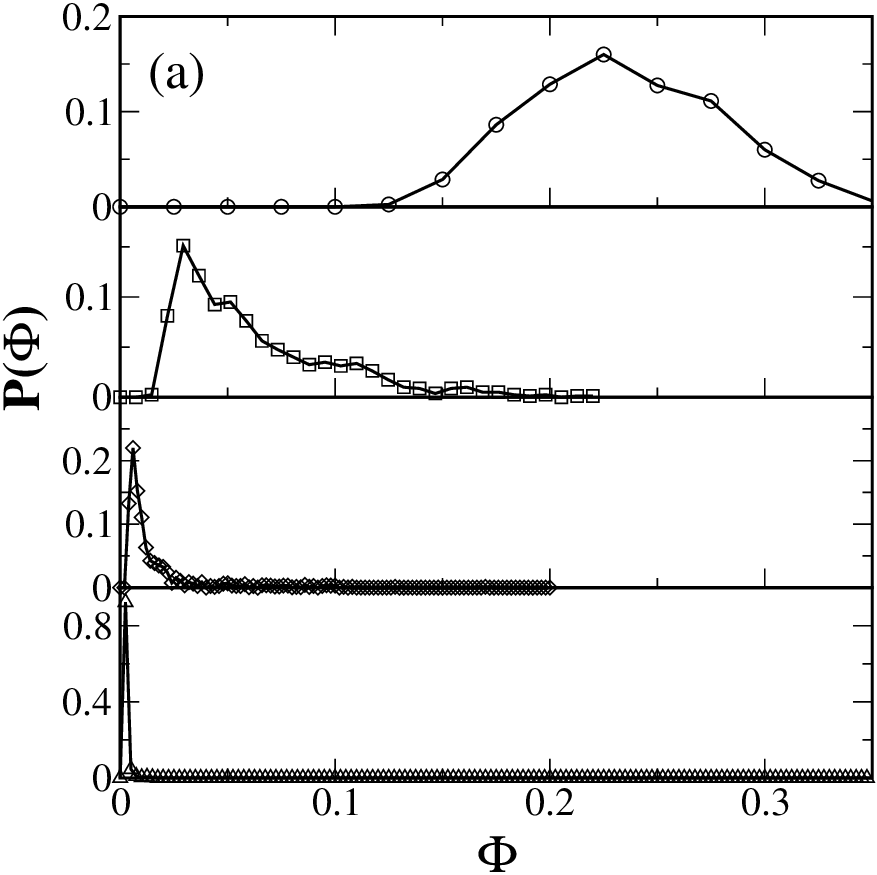}
\includegraphics[scale=0.4,clip]{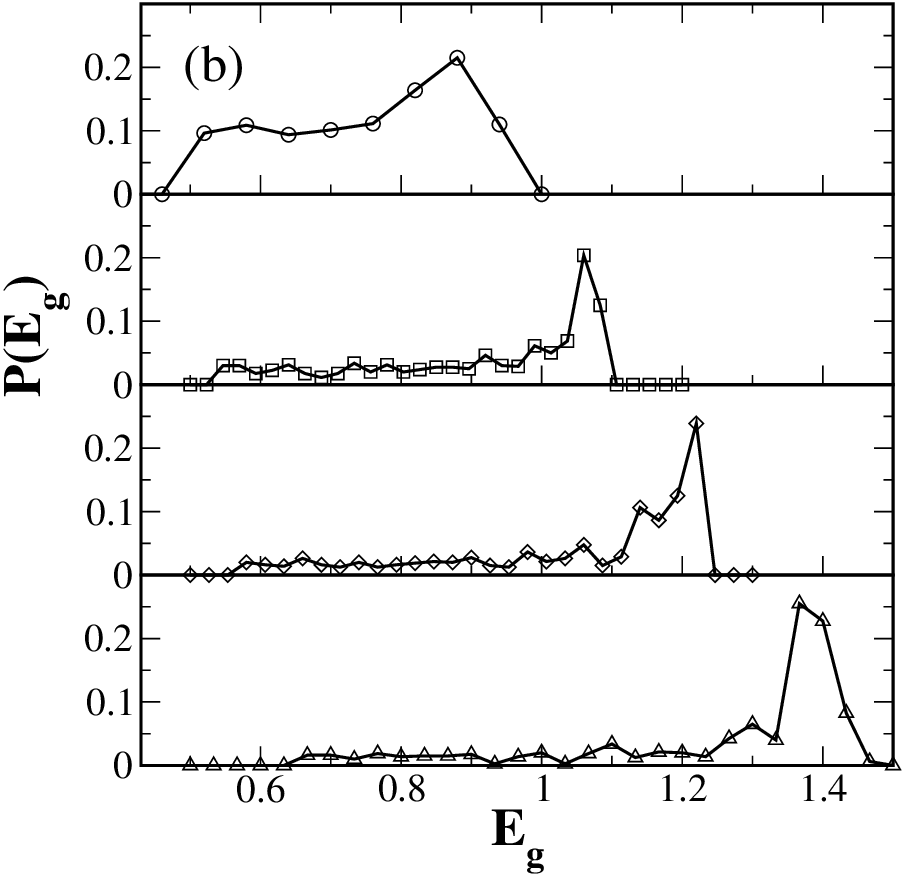}
\caption{Probability distribution of (a) local superconducting order parameter and
(b) spectral gap for different values of $x=0.5, 1.0, 1.5, 2.0$ (from top to bottom) at $U=2.0$.
}
\label{fig:fig5}
\end{figure}

In Fig.~\ref{fig:fig5}(a), the probability distribution function of local superconducting order parameter (PDF-OP)  for
 different values
of $x$ is shown. For small disorder, the PDF-OP is broad, and peaked moderately at a certain typical value of $\Phi$. With increasing $x$ 
the typical $\Phi$ decreases sharply and the PDF-OP narrows down considerably. This indicates that, while in the weak disorder
limit, arithmetic averaging may be used, in the strong disorder limit the typical value will manifest macroscopically. This is also reflected in the probability distribution function for spectral gap (PDF-SG). As expected, the PDF-SG is also broad at weak disorder and narrows considerably
at large disorder. In fact, since the weight contained in the peak is almost  more than 50\%, most of the sites
will have a gap in the neighbourhood of the gap value corresponding to the peak. Since the peak occurs at higher values of the gap
with increasing $x$, the gap in the CPA spectral function, shown in the bottom panel of Fig.~\ref{fig:fig3}, increases
with increasing $x$. In previous literature based on BdGMF or QMC of finite lattices incorporating inhomogeneous order
parameters, this increase in gap as a function of $x$ has been attributed to decrease in localization
length. Our results are based on the CPA, which is known to ignore localization effects. Thus we suggest that the
arguments based on PDF are sufficient, and localization physics need not be invoked to explain the increase of the insulating gap.
A more sophisticated treatment of this problem could be through typical medium theory, and subsequently statistical DMFT.
Although the latter has been carried out~\cite{Kawa}, this specific issue has not been addressed. 

\subsection{Fixed disorder; varying interaction strength}

\begin{figure}[ht]
\centering
\includegraphics[scale=0.4,clip]{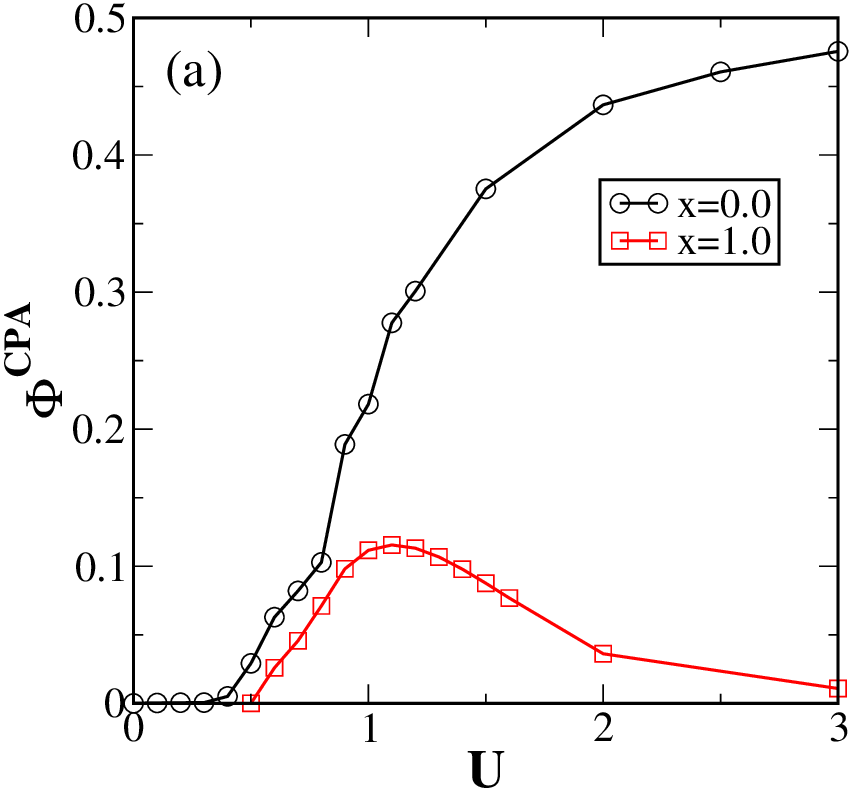}
\includegraphics[scale=0.4,clip]{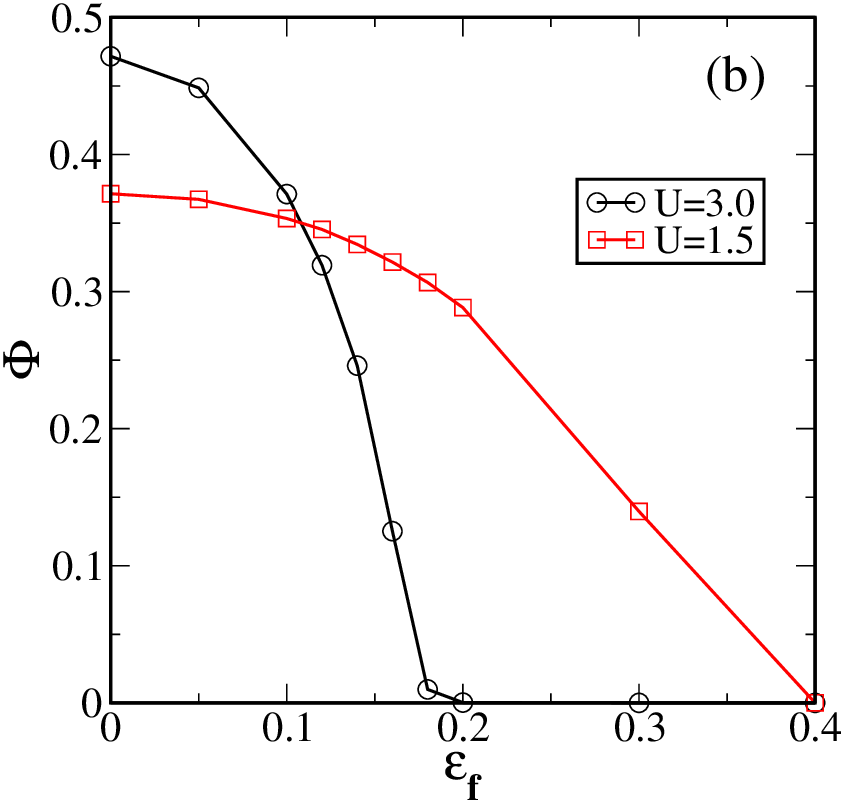}
\caption{(a) Superconducting order parameter of the disordered AHM as a function of $U$ at $x=0, 1.0$; (b) The $\Phi$ of the clean AHM as a function of site-energy $\epsilon_f$ for fixed $U=1.5$ and $3.0$.}
\label{fig:fig4}
\end{figure}

We have considered the behaviour of physical quantities as a function of 
disorder at a fixed interaction strength.  Now, we show the order parameter as a function
of $U$ at a fixed disorder. 
In the Fig.~\ref{fig:fig4}(a), the superconducting order parameter
computed at $x=0$ and $x=1.0$ is shown
as a function of $U$. For the clean case ($x=0$), as mentioned before, $\Phi$
increases and eventually saturates with increasing $U$. However, at finite 
disorder, $\Phi^{CPA}$ increases, reaches a maximum and subsequently
decreases.  Thus when we turn on disorder, the order parameter dependence
on $U$ changes qualitatively.   This can again be understood very simply  from the
clean AHM result, that at a fixed $U$, $\Phi$ decreases with increasing particle-hole asymmetry
defined by $\eta=1-2|(\epsilon_i-\mu)/U|$, as shown in Fig.~\ref{fig:fig4}(b).
For a larger $U$, the decrease of $\Phi$ with increasing $\eta$ is much more
rapid. Increasing site-disorder implies creating a greater number of sites with
large $\eta$, which would have a smaller $\Phi$ as compared to the sites
with $\eta\sim 0$. Thus, with increasing interaction strength, and a fixed disorder
concentration, the superconducting order parameter should decrease, and is indeed obtained.

\subsection{`Phase diagram'}

\begin{figure}[ht]
\centering
\includegraphics[scale=0.4,clip]{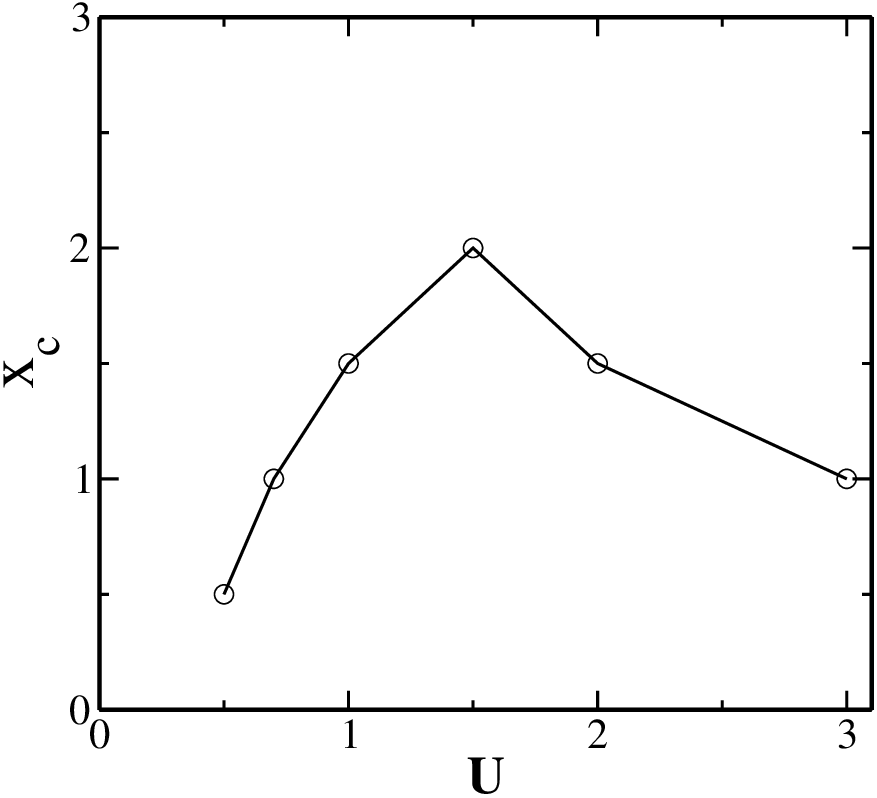}
\caption{The critical disorder $x_c(U)$ beyond which the superconducting state is completely destroyed is shown as a function of $U$.}
\label{fig:fig13}
\end{figure}
The variation of the spectral gap and the superconducting order parameter, shown in Fig.~\ref{fig:fig3}, was for a specific
interaction strength, namely $U=2.0$. We have repeated our calculations for various other $U$ values,
and could thus find the critical disorder strength $x_c(U)$, beyond which
the superconducting state is completely destroyed.
As shown in Fig.~\ref{fig:fig13}, we find that for large $U$, the $x_c(U)$ decreases with increasing the interaction strength.
 This kind of behaviour is also seen in QMC calculations of finite lattices through extrapolation~\cite{Nandini}. The
 superconducting state persists to higher disorder values with increasing $U$ in the weak to moderate coupling regime, in contrast to the strong coupling regime.

\section{Conclusions}
In this paper, we have studied the effects of site-disorder on a s-wave superconducting 
state as represented by an inhomogeneous attractive Hubbard model. Our
theoretical approach combines DMFT with IPTSC as an impurity solver and the CPA. 
Detailed studies of (a) the clean system away from half-filling and (b) probability
 distributions of the spectral gap and order parameter have been carried out. We have computed single-particle quantities such as the diagonal and off-diagonal spectral
functions in the disorder-$U$ plane. Using these, we obtained the spectral gap, superconducting order parameter and their probability distributions for different values of disorder and interaction strength. Some of our results agree qualitatively with those of 
previous studies~\cite{Amit,Sanjeev,Nandini}.  These include the non-monotonic dependence of the spectral
gap on the interaction strength, the destruction of the superconducting state with
disorder and a concomitant superconductor-insulator transition. These studies~\cite{Amit,Sanjeev,Nandini} were carried out on two-dimensional lattices, while our work is within DMFT. Thus, we conclude that
dimensionality has little role to play in these results.  We further argue that our results may be explained by utilizing
 particle-hole asymmetry and disorder induced probability distributions, with no necessity
to invoke localization physics. 

    In order to probe localization physics within local approaches, one must ideally incorporate short-range correlations using techniques such as typical medium-dynamical cluster approximation~\cite{Mark}.
The extension of our results to finite temperature and general fillings would pave the way to comparison with experiments. 

\section{Acknowledgments}
The authors thank CSIR, India and JNCASR, India for funding this research.


\begin{thebibliography}{100}
\bibitem{PA} P.\ A.\ Lee and T.\ V.\ Ramakrishnan, Rev.\ Mod.\ Phys.\ {\bf 57},
287 (1985).
\bibitem{Belitz} D.\ Belitz and T.\ R.\ Kirkpatrick, Rev.\ Mod.\ Phys.\ {\bf 66}, 261
(1994).
\bibitem{Pratap1} Mintu Mondal {\it et al}, Phys.\ Rev.\ Lett.\ {\bf 106}, 047001 (2011).
\bibitem{Pratap2} Madhavi Chand {\it et al}, Phys.\ Rev.\ B {\bf 85}, 014508 (2012).
\bibitem{Crane} R.\ Crane, N.\ P.\ Armitage, A.\ Johansson, G.\ Sambandamurthy, 
D.\ Shahar, G.\ Gruner,  Phys.\ Rev.\ B {\bf 75}, 184530 (2007͒).
\bibitem{Arti} A.\ Garg, H.\ R.\ Krishnamurthy, and M.\ Randeria, Phys.\ Rev.\ B
{\bf 72}, 024517 (2005).
\bibitem{Bauer} J.\ Bauer, A.\ C.\ Hewson, and N.\ Dupuis, Phys.\ Rev.\ B {\bf 79}, 214518 (2009).
\bibitem{Werner} Akihisa Koga, Philipp Werner, Phys.\ Rev.\ A {\bf 84}, 023638 (2011).
\bibitem {Antoine} A.\ Georges, G.\ Kotliar, W.\ Krauth, and M.\ J.\ Rozenberg, 
Rev.\ Mod.\ Phys.\ {\bf 68}, 13 (1996).
\bibitem{Kotliar} G.\ Kotliar, D.\ Vollhardt, Physics Today, (2004).
\bibitem{Dieter} D.\ Vollhardt, AIP Conference Proceedings {\bf 1297}, 339 (2010) .
\bibitem{Kawa}  Masaru Sakaida, Kazuto Noda, and Norio Kawakami, 
J.\ Phys.\ Soc.\ Japan {\bf 82} 074715 (2013).
\bibitem{Anderson} P.\ W.\ Anderson, J.\ Phys.\ Chem.\ Solids {\bf 11}, 26 (1959).
\bibitem{Gorkov} A.\ A.\ Abrikosov and L.\ P.\ Gorkov, Sov.\ Phys.\ JETP {\bf 9},
 220 (1959).
\bibitem{Amit} A.\ Ghosal, M.\ Randeria and N.\ Trivedi, Phys.\ Rev.\ B {\bf 65},
014501 (2001).
\bibitem{Sanjeev} Sanjeev Kumar and Prabuddha B.\ Chakraborty, arxiv:1302.1967.
\bibitem{Nandini} R.\ T.\ Scalettar, N.\ Trivedi, C.\ Huscroft, Phys.\ Rev.\ B {\bf 59},
 4364 (1999).
\bibitem{Elliott} R.\ J.\ Elliott, J.\ A.\ Krumhansl, and P.\ L.\ Leath, Rev.\ Mod.\ Phys.\
{\bf 46}, 465 (1974).
\bibitem{Jannis} V.\ Janis and D.\ Vollhardt, Phys.\ Rev.\ B {\bf 46}, 15712 (1992).
\bibitem{pott} M.\ Potthoff and M.\ Balzer, \prb {\bf 75}, 125112 (2007).
\bibitem{Mark} Chinedu E.\ Ekuma {\it et al}, arXiv:1306.5712.
\end{thebibliography}
\end{document}